\documentclass[12pt]{article}
\usepackage{amssymb}
\usepackage{epsfig,caption,graphicx,eepic,epic}
\usepackage{pstricks}
\usepackage{amsmath}
\usepackage{multido}

\begin{document}
\def\be{\begin{equation}}
\def\ee{\end{equation}}
\def\ba{\begin{eqnarray}} 
\def\ea{\end{eqnarray}}
\def\nn{\nonumber}

\newcommand{\bbf}{\mathbf}
\newcommand{\rrm}{\mathrm}

\title{Application of a renormalization algorithm in Hilbert space to the study of 
many-body quantum systems\\} 

\author{Tarek Khalil
\footnote{E-mail address: khalil@lpt1.u-strasbg.fr}\\ 
and\\
Jean Richert
\footnote{E-mail address: richert@lpt1.u-strasbg.fr}\\ 
Laboratoire de Physique Th\'eorique, UMR 7085 CNRS/ULP,\\
Universit\'e Louis Pasteur, 67084 Strasbourg Cedex,\\ 
France} 
 
\date{\today}
\maketitle 
\begin{abstract}
We implement an algorithm which is aimed to reduce the number of basis states spanning 
the Hilbert space of quantum many-body systems. We test the procedure by  working out 
and analyzing the spectral properties of strongly correlated and frustrated quantum spin 
systems. The role and importance of symmetries are investigated.
\end{abstract} 
\maketitle

\section{Introduction.}

Most microscopic many-body quantum systems are subject to strong interactions which act 
between their constituents. The description of some systems can be tackled by means of 
perturbation theory. This may be the case when it is possible to introduce a mean-field concept
which is able to include quantitatively the main part of the interactions, leaving 
a remaining residual contribution which acts as a more or less small perturbation.\\

Often such an approach does not lead to sensible results, in particular when one is 
dealing with realistic quantum spin systems. Non-perturbative techniques are needed. During 
the last decades a considerable amount of procedures relying on the renormalization group 
concept introduced by Wilson ~\cite{wil} have been proposed and tested. Some of them are 
specifically devised for quantum spin systems, like the Real Space Renormalization Group
(RSRG)~\cite{mal,cap,whi} and the Density Matrix Renormalization Group (DMRG)
~\cite{malve,whi2,henk}.\\  

In many cases the study of the spectral properties of quantum systems is obtained through the 
diagonalization of a many-body Hamiltonian in Hilbert space spanned by a complete, in general 
infinite or at least very large set of basis states although the information of interest is 
restricted to the knowledge of a few low energy states generally characterized by collective 
properties. Consequently it is necessary to manipulate very large matrices in order to extract 
a reduced quantity of informations. \\

Recently we proposed a non-perturbative approach which tackles this question ~\cite{khri}. 
The procedure consists of an algorithm which implements a step by step reduction of the 
size of Hilbert space by means of a projection technique. It relies on the renormalization 
concept following in spirit former work based on this concept~\cite{gla,mue,bek}.
Since the reduction procedure does not act in ordinary or momentum space but in the zero-
dimensional Hilbert space like in the procedure developed in ref.~\cite{bae}, it is in principle 
applicable to all types of microscopic quantum systems, in contradistinction with methods like the 
DMRG procedure which works in ordinary space and is specifically applicable to $1d$ quantum 
lattice systems.\\
 
In the present work we implement this algorithm as a preliminary test of the practical efficiency 
and accuracy of the method when applied to strongly interacting systems which cannot be 
treated by means of perturbative methods. Second we want to see how far it is able to deliver 
physical information about the properties of the many-body systems it is aimed to describe.
We use quantum spin systems as first test probes.\\

The outline of the paper is the following. In section $2$ we recall the essential steps leading 
to the derivation of the equation which governs the evolution of the coupling strength attached 
to the interaction. The reduction formalism is universal in the sense that it works for any kind of 
many-body quantum system. Section $3$ is devoted to the application of the algorithm to 
frustrated quantum spin ladders with two legs and one spin per site. We analyze the outcome of the
applied algorithm on systems of different sizes and characterized by different coupling strengths
by means of numerical examples, with bases of states developed in the $SU(2)$ and $SO(4)$-symmetry 
scheme. General conclusions and further planned investigations and developments are drawn in 
section $4$.\\

\section{The reduction algorithm.}

\subsection{General concept: the space reduction procedure.}
  
We consider a system of quantum objects (particles, spins) which are characterized by a
discrete spectrum. The system is governed by a Hamiltonian 
$H^{(N)}\left ( g_1^{(N)},g_2^{(N)},...,g_p^{(N)}\right)$ which depends on $p$ coupling
strengths\\ 
$\left\{ g_1^{(N)},g_2^{(N)},\cdots, g_p^{(N)}\mapsto g^{(N)}\right\}$ 
and acts in a Hilbert space $ {\cal H}^{(N)}$ of dimension $N$. The spectrum is obtained
from  
\be
H^{(N)}(g^{(N)}) |\Psi_i^{(N)}(g^{(N)})\rangle = \lambda_i(g^{(N)}) 
|\Psi_i^{(N)}(g^{(N)})\rangle  
\label{eq0} \ .
\ee
where the eigenvalues $\left\{\lambda_i(g^{(N)}) , \, i=1,\cdots, N\right\}$
and the eigenstates 

$\left\{|\Psi_i^{(N)}(g^{(N)})\rangle,\, i=1,\cdots,N\right\}$
depend on the set of coupling constants $\{g^{(N)}\}$. If the relevant quantities of interest 
are for instance $M$ eigenvalues out of the original set it makes sense to try to
define a new effective Hamiltonian $H^{(M)}(g^{(M)})$ whose eigenvalues reproduce 
the $M$ selected states and verifies
\be
H^{(M)}(g^{(M)}) |\Psi_i^{(M)}(g^{(M)})\rangle = \lambda_i(g^{(M)}) 
|\Psi_i^{(M)}(g^{(M)})\rangle  
\label{eq1} \ 
\ee
with the constraints
\be
\lambda_i(g^{(M)}) = \lambda_i(g^{(N)})
\label{eq2} \ 
\ee
for $i = 1,...,M$. If this can be realized Eq. $(3)$ implies a relation between the 
coupling constants in the original and reduced space
\be
g_k ^{(M)} = f_k(g_1^{(N)},g_2^{(N)},...,g_p^{(N)})
\label{eq3} \ 
\ee
with $k = 1,...,p$. The effective Hamiltonian $H^{(M)}(g^{(M)})$ may not be rigorously 
derivable from $H^{(N)}$. It should be constructed so that it optimizes the overlap between 
the original and reduced set of eigenstates. We show next how this space reduction may be 
implemented in practice.\\ 
 
\subsection{Reduction procedure and renormalization of the coupling strengths.}

We sketch the procedure which leads from Eq.~(\ref{eq0}) to Eq.~(\ref{eq1}). Details can be found
in ref.~\cite{khri}.\\

We consider a system described by a Hamiltonian depending on a unique coupling strength $g$
which can be written as a sum of two terms 
\be
H = H_0 + g H_1 
\label{eq4} \  
\ee

The Hilbert space  ${\cal H}^{(N)}$ of dimension $N$ is spanned by an a priori arbitrary
set of basis states 
$\left\{|\Phi_i\rangle, \,i=1,\cdots, N\right\}$. An eigenvector $|\Psi_1^{(N)}\rangle$
can be written as
\be
|\Psi_1^{(N)}\rangle = \sum_{i=1}^{N}  a_{1i}^{(N)}(g^{(N)})|\Phi_i\rangle
\label{eq5} \  
\ee    
where the amplitudes $\{a_{1i}^{(N)}(g^{(N)})\}$ depend on the value $g^{(N)}$ of
$g$ in  ${\cal H}^{(N)}$.\\

Using the Feshbach formalism~\cite{fesh} the Hilbert space may be decomposed into subspaces by 
means of the projection operators $P$ and $Q$, 
\be
{\cal H}^{(N)} = P{\cal H}^{(N)} + Q{\cal H}^{(N)}
\ee

In practice the subspace $ P{\cal H}^{(N)}$ is chosen to be of dimension 
$\mathrm{dim}\,P{\cal H}^{(N)}= N-1$ by elimination of 
one  basis state. The projected eigenvector $P|\Psi_1^{(N)}\rangle$ obeys the Schro\"edinger
equation  
\be
H_{eff}(\lambda_1^{(N)})P |\Psi_1^{(N)}\rangle =  \lambda_1^{(N)}
P |\Psi_1^{(N)}\rangle
\label{eq6} \ . 
\ee
where $H_{eff}(\lambda_1^{(N)})$ is the effective Hamiltonian which operates in the subspace  
$P{\cal H}^{(N)}$. It depends on the eigenvalue $\lambda_1^{(N)}$ which is the eigenenergy 
corresponding to $|\Psi_1^{(N)}\rangle$ in the initial space ${\cal H}^{(N)}$. The coupling 
$g^{(N)}$ which characterizes the Hamiltonian $H^{(N)}$ in ${\cal H}^{(N)}$ is now aimed to 
be changed into $g^{(N-1)}$ in such a way that the eigenvalue in the new space 
${\cal H}^{(N-1)}$ is the same as the one in the complete space
\be
\lambda_1^{(N-1)} = \lambda_1^{(N)} 
\label{eq7} \  
\ee
The determination of $g^{(N-1)}$ by means of the constraint expressed by Eq.~(\ref{eq7}) 
is the central point of the procedure. In practice the reduction of the vector space
from $N$ to $N-1$ results in a renormalization of the coupling constant from $g^{(N)}$ 
to $g^{(N-1)}$ preserving the physical eigenenergy $\lambda_1^{(N)}$.\\

In the sequel $P|\Psi_1^{(N)}\rangle$ is chosen to be the ground state eigenvector and 
$\lambda_1^{(N)} = \lambda_1^{(N-1)} = \lambda_1$ the corresponding eigenenergy. In 
ref.~\cite{khri} it is shown how $g^{(N-1)}$ can be obtained as a solution of an algebraic 
equation of the second degree, see Eq. (17). The reduction procedure is iterated in a step 
by step decrease of the dimensions of the vector space, $N \mapsto N-1 \mapsto N-2 \mapsto...$ 
leading at each step $k$ to a coupling strength $g^{(N-k)}$ which can be given as the solution 
of a flow equation in a continuum limit description of the Hilbert space ~\cite{khri,jr}. 
The procedure can be generalized to Hamiltonians depending on several coupling constants.\\

\subsection{Reduction algorithm.}

$1-$ Consider a quantum system described by an Hamiltonian $H^{(N)}$ which acts in an 
$N$-dimensional Hilbert space.
\\
$2-$ Compute the matrix elements of the Hamiltonian matrix $H^{(N)}$ in a definite basis of 
states $\{|\Phi_i\rangle, i=1,\ldots,N \}$. The diagonal matrix elements 
$\{\epsilon_i = \langle \Phi_i|H^{N}| \Phi_i \rangle\}$ are arranged in increasing order with 
respect to the $\{\epsilon_i\}$.\\
$3-$ Use the Lanczos technique to determine $\lambda_1^{(N)}$ and $|\Psi_1^{(N)}(g^{(N)})\rangle$.
$\lambda_1^{(N)}$ may be chosen as the experimental value $\lambda_1$ if known.\\
$4-$ Fix $g^{(N-1)}$ as described in section 2.2. Take the solution of the algebraic second 
order equation closest to  $g^{(N)}$ (Eq. (17) in ~\cite{khri}).\\
$5-$ Construct $H^{(N-1)} = H_0 + g^{(N-1)} H_1$ by elimination of the matrix elements of 
$H^{(N)}$ involving the state $|\Phi_N\rangle$.\\
$6-$ Repeat the procedures $2$, $3$, $4$ and $5$ by fixing at each step $k$
$\lambda_1^{(N-k)}=\lambda_1^{(N)} = \lambda_1$.\\ 
$7-$ The iterations may be stopped at $N = N_{min}$ corresponding to the limit of space dimensions for which the spectrum gets unstable.

\subsection{Preliminary remarks.}

\begin{itemize}

\item The procedure is aimed to generate the energies and other physical properties of the ground 
state and low-energy excited states of strongly interacting systems.

\item The implementation of the reduction procedure asks for the knowledge of $\lambda_1$ and 
the corresponding eigenvector $|\Psi_1^{(N-k)}\rangle$ at each step $k$ of the reduction 
process. The eigenvalue $\lambda_1$ is chosen as the physical ground state energy of the 
system. Eigenvalue and eigenvector can be obtained by means of the Lanczos algorithm
~\cite{lanc1,lanc2,henk} which is particularly well adapted to very large vector space dimensions. 
This algorithm is used here in order to fix $\lambda_1$ and $|\Psi_1^{(N-k)}\rangle$.\\ 

\item The process does not guarantee a rigorous stability of the eigenvalue $\lambda_1$. Indeed one 
notices that $|\Psi_1^{(N-k-1)}\rangle$ which is the eigenvector in the space ${\cal H}^{(N-k-1)}$ 
and the projected state $P|\Psi_1^{(N-k)}\rangle$ of $|\Psi_1^{(N-k)}\rangle$ into 
${\cal H}^{(N-k-1)}$ may differ from each other. As a consequence it may not be possible to keep  
$\lambda_1^{(k-1)}$ rigorously equal to $\lambda_1^{(k)} = \lambda_1$. In practice the degree 
of accuracy depends on the relative size of the eliminated amplitudes $\{a_{1k}^{(k)}(g^{(k)})\}$.
This point will be tested by means of numerical estimations and further discussed below.\\  

\item The algorithm and different points which have been quoted above will now be developed in 
applications of the procedure to explicit models, here frustrated spin ladders. The applicability 
of the algorithm will be tested in different symmetry schemes.
   
\item The Hamiltonians of the considered ladder systems are characterized by a fixed total magnetic 
magnetization $M_{tot}$. We shall work in subspaces which correspond to fixed $M_{tot}$. The total
spin $S_{tot}$ is also a good quantum number which defines smaller subspaces for fixed $M_{tot}$. 
We shall not introduce them here because projection procedures on $S_{tot}$ are time consuming.
Furthermore we want to test the algorithm in large enough spaces although not necessarily the largest
possible ones in this preliminary tests considered here. 

\end{itemize}

\section{Application to frustrated two-leg quantum spin ladders.}

\subsection{The model}

\subsubsection{SU(2)-symmetry framework.}

Consider spin-$1/2$ ladders~\cite{lin1,lin2} described by Hamiltonians of the following type 
and shown in Fig. 1.
\ba\label{eq8}
H^{(s,s)} &=& J_t\sum_{i=1}^{L} s_{i_1}s_{i_2} + J_l\sum_{<ij>}s_{i_1}s_{j_1} +
J_l\sum_{<ij>}s_{i_2}s_{j_2} + J_{1c}\sum_{(ij)}s_{i_1}s_{j_2}  
\\ \nn
&  & + J_{2c}\sum_{(ij)}s_{i_2}s_{j_1} 
\ea
 
\begin{figure}
\epsfig{file=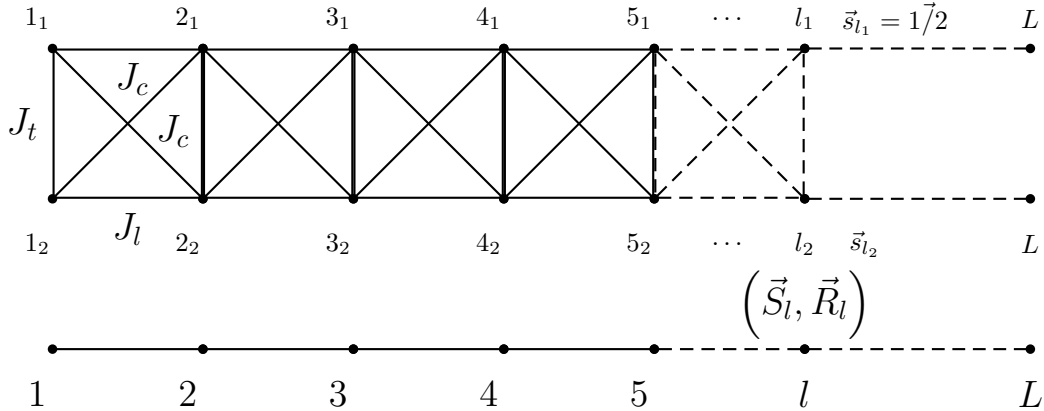}
\caption{Top: the original spin ladder. The coupling strengths are indicated as given in the text. 
Bottom: The ladder in the SO(4) representation. See the text.}
\end{figure}

The indices $1$ or $2$ label the spin $1/2$ vector operators $s_{i_k}$ acting on the sites $i$ 
on both ends of a rung, in the second and third term $i$ and $j$ label nearest neighbours, here 
$j = i + 1$ along the legs of the ladder. The fourth and fifth term correspond to diagonal 
interactions between sites located on different legs, $j = i + 1$. $L$ is the number of sites on 
a leg (Fig. 1) where $J_{1c} = J_{2c} = J_c$. The coupling strengths $J_t, J_l, J_{c}$ are 
positive.\\

As stated above the renormalization is restricted to a unique coupling strength, see 
Eq.~(\ref{eq4}). It is implemented here by putting $H_0 = 0$ and $H^{(N)} = g^{(N)} H_1$ where 
$g^{(N)} = J_t$ and
\be
H_1 = \sum_{i=1}^{L} s_{i_1}s_{i_2} + \gamma_{tl} \sum_{<ij>}(s_{i_1}s_{j_1} +
s_{i_2}s_{j_2})  + \gamma_{c}\sum_{<ij>}(s_{i_1}s_{j_2} + s_{i_2}s_{j_1}) 
\label{eq9} \ . 
\ee
where $\gamma_{tl} = J_{l}/J_{t}$, $\gamma_{c} = J_{c}/J_{t}$. These quantities are kept constant 
and $g^{(N)} = J_t$  will be subject to renormalizationin the reduction process.\\

The basis of states $\{|\Phi_k\rangle\, k=1,\ldots,N\}$ is chosen as

\be\nonumber
|\Phi_k\rangle = |1/2 ~~m_1,...,1/2 ~~m_i,...,1/2 ~~m_{2L}, \sum_{i=1}^{2L} m_i= M_{tot} = 0\rangle  
\ee
with $\{m_i = +1/2, -1/2\}$.

\subsubsection{SO(4)-symmetry framework.}

The basis of states may be written in an $SO(4)$-symmetry scheme. Different choices 
of bases may induce a more or less efficient reduction procedure depending on the strength 
of the coupling constants $J_{t}, J_{l}, J_{c}$. This point is investigated here. 

By means of a spin rotation ~\cite{kika,kimo}
\be
 s_{i_1} = \frac{1}{2} (S_i + R_i)
\label{eq10} \ .
\ee
\be
 s_{i_2} = \frac{1}{2} (S_i - R_i)
\label{eq11} \ .
\ee
the Hamiltonian Eq.(10) can be expressed in the form\\
 
\be
H^{(S,R)} = \frac{J_t}{4}\sum_{i=1}^{L} (S_{i}^{2} -  R_{i}^{2}) + J_1\sum_{<ij>}S_i S_j + 
J_2\sum_{<ij>}R_i R_j
\label{eq12} \   
\ee
The structure of the corresponding system is shown in the lower part of Fig. 1.
Here  $J_{1}=(J_{l}+J_{c})/2$, $J_{2}=(J_{l}-J_{c})/2$  and as before $J_{1c}=J_{2c}=J_{c}$.
The components $S_{i}^{(+)}, S_{i}^{(-)}, S_{i}^{(z)}$ and $R_{{i}}^{(+)}, R_{{i}}^{(-)},
R_{{i}}^{(z)} $ of the vector operators $S_{i}$ and  $R_{i}$ are the $SO(4)$ group generators 
and $<ij>$ denotes nearest neighbour indices
 
\ba\nonumber
S_{i}^{(+)}  = \sqrt{2}(X^{(11)(10)}_{i} + X^{(10)(1-1)}_{i}) = S_{i}^{(-)*}
\ea
 
\ba\nonumber
S_{i}^{(z)}  =  X^{(11)(11)}_{i} - X^{(1-1)(1-1)}_{i}
\ea

\ba\nonumber
R_{i}^{(+)}  =  \sqrt{2}(X^{(11)(00)}_{i} - X^{(00)(1-1)}_{i}) = R_{i}^{(-)*}
\ea
 
\ba\nonumber
R_{i}^{(z)} =  - (X^{(10)(00)}_{i} + X^{(00)(10)}_{i})
\ea 

where 
\ba\nonumber
X^{(S_i M_i)(S'_i  M'_i)}_{i} = |S_i M_i \rangle  \langle S'_i  M'_i|\nonumber
\ea

In this framework the states $\{|S_{i}M_{i}\rangle\}$ are defined as  
\ba\nonumber
|S_i M_i \rangle = \sum_{m_1,m_2} \langle 1/2 ~~m_1 ~~1/2 ~~m_2|S_i M_i  \rangle
|1/2 ~~m_1\rangle_{i}  |1/2 ~~m_2\rangle_{i}
\ea
along a rung are coupled to $S_i = 0$ or $S_i = 1$. Spectra are constructed in this representation 
as well as in the $SU(2)$ representation and the basis of states  $\{|\Phi_k\rangle\}$
takes the form

\be\nonumber
|\Phi_k\rangle = |S_1 M_1,...,S_i M_i,...,S_{L} M_{L}, \sum_{i=1}^{L} M_i= M_{tot} = 0\rangle  
\ee

\subsection{Test observables}
In order to quantify the accuracy of the procedure we introduce different test quantities in order 
to estimate quantitatively deviations between ground state and low excited state energies in 
Hilbert spaces of different dimensions. The stability of low-lying states can be estimated by means 
of    

\ba\label{13}
p(i) = |\frac{(e_i^{(N)}-e_i^{(N-k)})}{e_i^{(N)}}| \times 100 & with& i=1,\ldots,4
\ea  
where $e_i^{(N-k)} = \lambda_i^{(N-k)}/2L$ corresponds to the energy per site at the $i$th physical 
state starting from the ground state at the $k$th iteration in Hilbert space. These quantities 
provide a percentage of loss of accuracy of the eigenenergies in the different reduced spaces.\\

A global characterization of the ground state wavefunction in different symmetry schemes can also 
be given by the entropy per site in a space of dimension $n$

\ba\label{14}
s = - \frac{1}{2L} \sum_{i=1}^{n}{P_i}ln{P_i} & with & P_i = |\langle\Phi_i^{(n)}|\Psi_1^{(n)}
\rangle|^{2} = |a_{1i}^{(n)}|^{2}
\ea
which works as a global measure of the distribution of the amplitudes $\{a_{1i}^{(n)}\}$ in the 
physical ground state.\\

\subsection{Spectra in the SU(2)-symmetry framework.}

We apply the reduction algorithm to ladders with two legs, different numbers of sites and 
different values of the coupling strengths. Results obtained with an $SU(2)$-symmetry basis
of states are shown in Figs. (2 - 4) and Fig. 6.\\ 

\subsubsection{First case: $L$= 6, $J_t$=15, $J_l$=5, $J_{c}$=3}
 
We choose the basis states in the framework of the $M$-scheme corresponding to subspaces with 
fixed values of the projection of the spin of the $\{|\Phi_i\rangle\}$, $M_{tot}=0$.\\ 

In the present case $J_t > J_l, J_{c}$.   The dimension of the subspace is reduced
step by step as explained above starting from $N=924$. As stated in section 2.3 the basis states  
$\{|\Phi_i\rangle\}$ are ordered with increasing energy of their diagonal matrix elements 
$\{\epsilon_i\}$ and eliminated starting from the state with largest energy, $\epsilon_N$.\\ 

\begin{figure}[ht]
\includegraphics{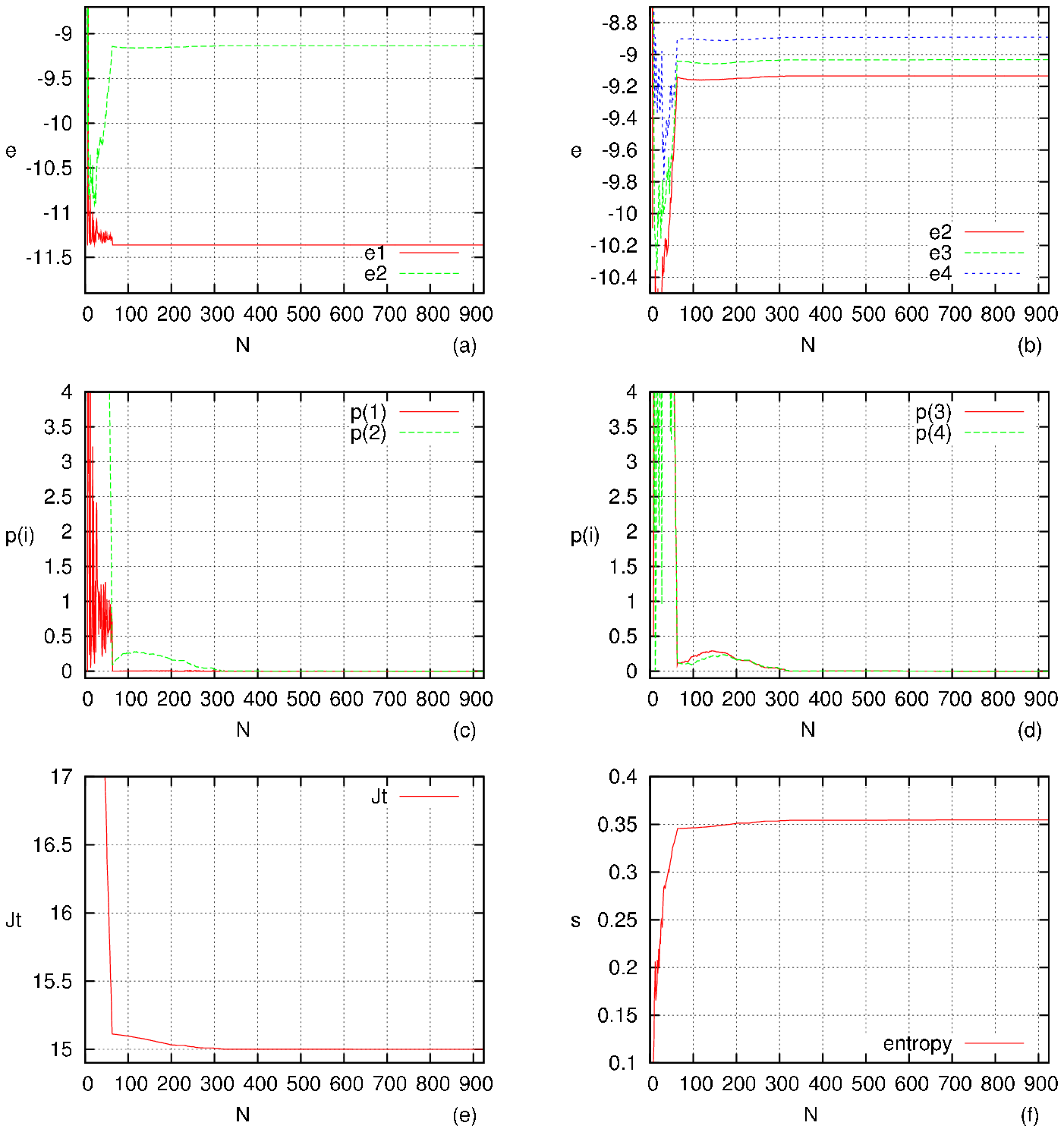}
\caption{$SU(2)-scheme$. $N$ is Hilbert space dimension. The $\{e_i, i= 1,2,3,4\}$ are the 
energies of the ground and excited states per site. $L=6$ sites along a leg. 
$J_t=15$, $J_l=5$, $J_{c}=3$}
\label{fig2}
\end{figure}

As seen in Fig.(\ref{fig2}a) the ground state of the system stays stable down to $n\sim50$
where $n$ is the dimension of the reduced space. The coupling constant $J_t$  does not move either 
up to $n\sim 300$. Figs.(\ref{fig2}a-b) show the evolution of the first excited states which 
follows the same trend as the ground state. Deviations from their initial value at 
$N=924$ can be seen in Fig.(\ref{fig2}c-d) where the $p(i)$'s defined above represent these 
deviations in terms of percentages.\\ 
 
For $n \leq 50$ the spectrum gets unstable, the renormalization of the coupling constant can 
no longer correct for the energy of the lowest state. Indeed the coupling constant $J_t$ increases
drastically as seen in Fig.(\ref{fig2}e). The reason for this behaviour can be found in 
the fact that at this stage the algorithm eliminates states which have an essential component in the 
state of lowest energy. The same message can be read on Fig.(\ref{fig2}f), the drop in the entropy  
per site $s$ is due to the elimination of sizable amplitudes $\{a_{1i}\}$.\\

\subsubsection{Second case: $L$= 6, $J_t$=5.5, $J_l$=5, $J_{c}$=3}
 
\begin{figure}[ht]
\includegraphics{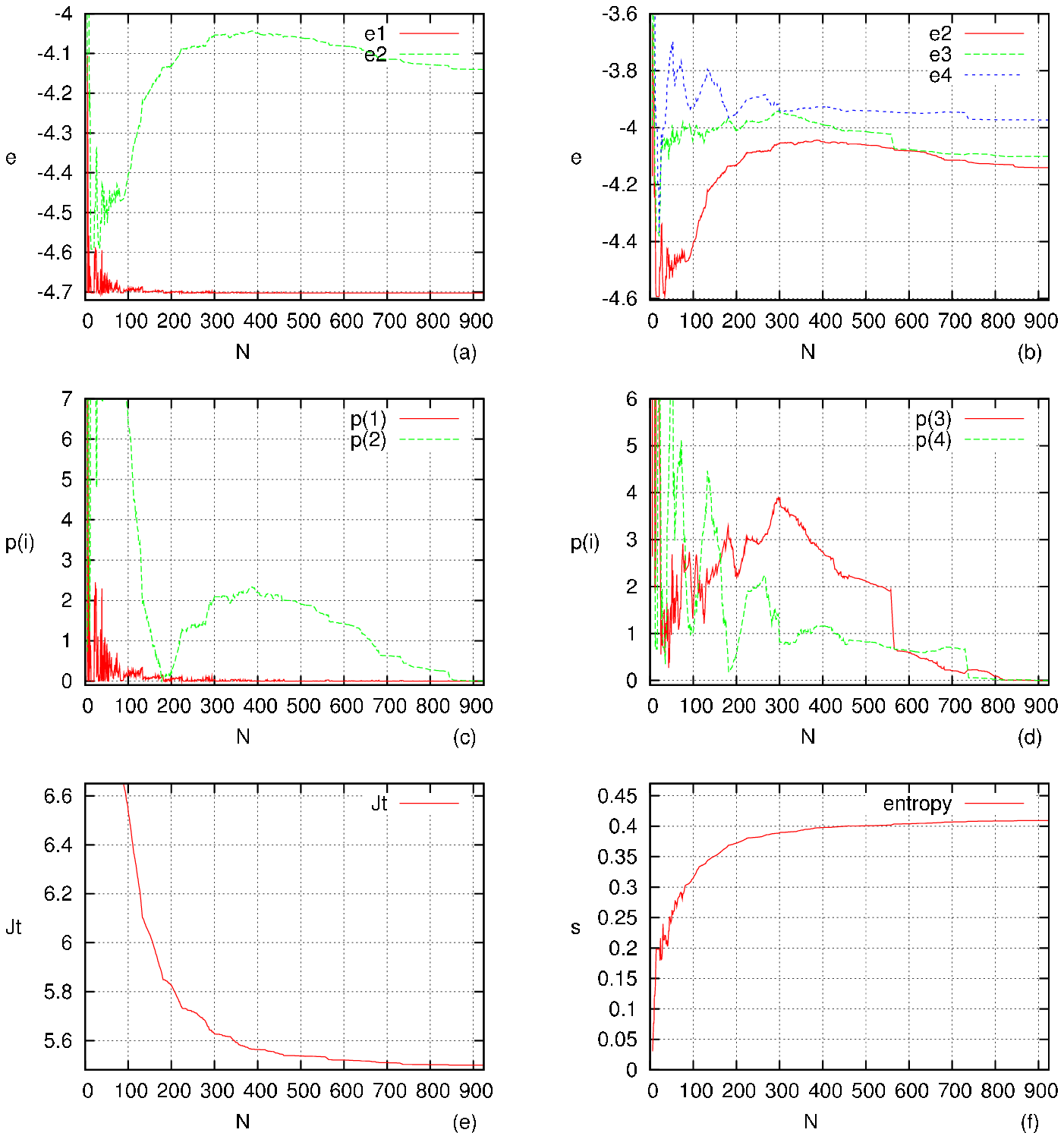}
\caption{$SU(2)- scheme$. $N$ is Hilbert space dimension. The  $\{e_i, i= 1,2,3,4\}$  are the 
energies of the ground and excited states per site. $L=6$ sites along a leg.
$J_t=5.5$, $J_l=5$, $J_{c}=3$}
\label{fig3}
\end{figure}

Contrary to the former case the coupling constant $J_t$ along rungs is now of the same strength as 
$J_l, J_{c}$. Results are shown in Fig.(\ref{fig3}). The lowest energy state is now 
stable down to $n \sim 100$. This is also reflected in the behaviour of the excited states which move
appreciably for $n \leq 200$. Fig.(\ref{fig3}e) shows that the coupling constant $J_t$ starts to 
increase sharply between $n=300$ and $n=200$. It is able to stabilize the excited states up to about 
$n=200$ and the ground state up to $n=70$. The instability for $n\leq70$ reflects in the evolution of
the $p(i)$'s, Figs.(\ref{fig3}c-d) which get of the order of a few percents. The entropy 
Fig.(\ref{fig3}f) follows the same trend.\\

Comparing the two cases above and particularly the entropies Fig.(\ref{fig3}f) and Fig.(\ref{fig2}f)
one sees that the stronger $J_t$ the more the amplitude strength of the ground state wavefunction 
is concentrated in a smaller number of basis state components. The elimination of sizable components
of the wavefunction leads to deviations which can be controlled up to a certain limit by means of 
the renormalization of $J_t$. One sees that large values of $J_t$ favour a low number of 
significative components in the low energy part of the spectrum in a $SU(2)$ symmetry framework.\\ 

\begin{figure}[ht]
\includegraphics{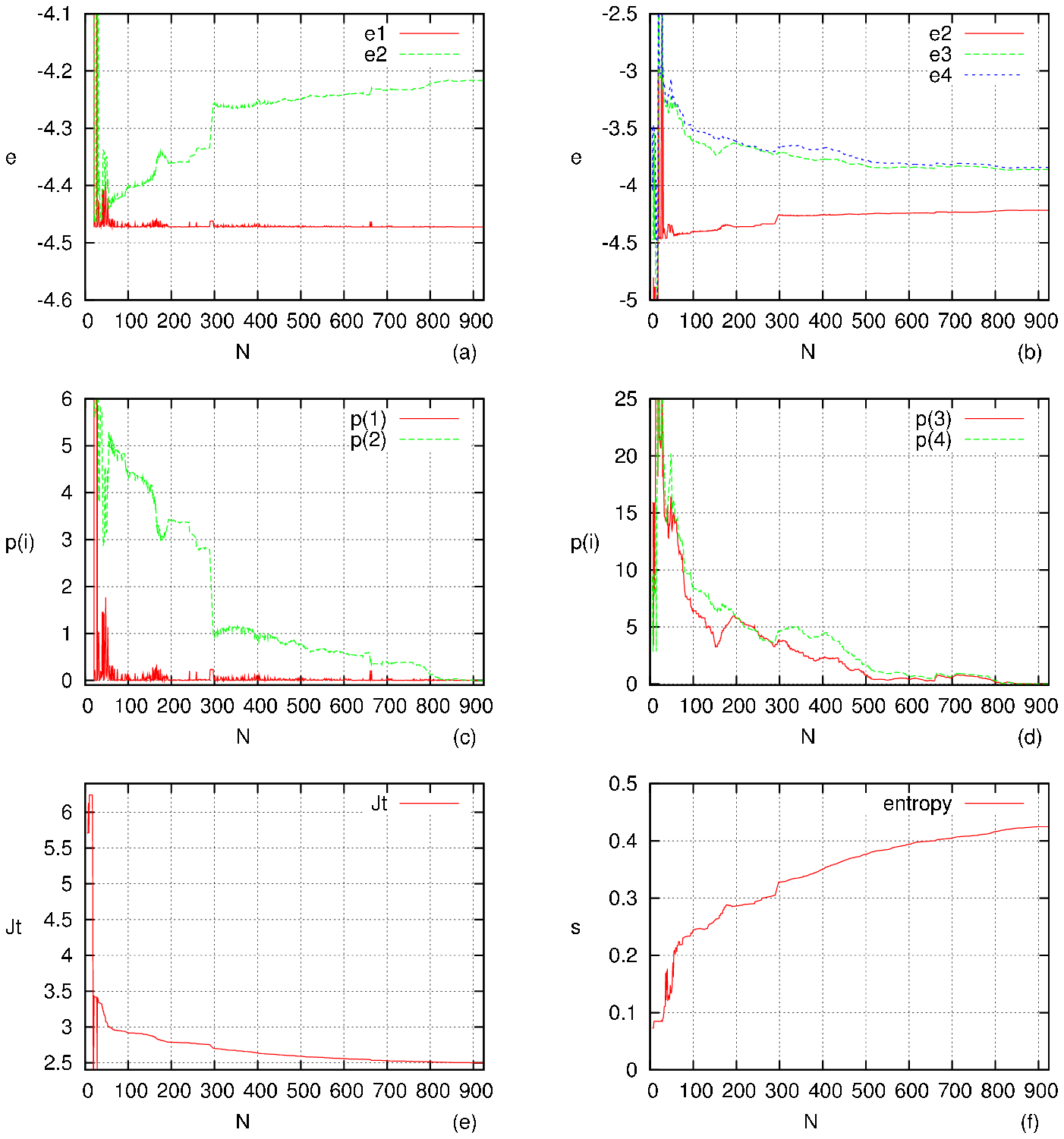}
\caption{$SU(2)- scheme$. $N$ is Hilbert space dimension. The  $\{e_i, i= 1,2,3,4\}$ are the 
energies per site. $L=6$ sites along a leg. $J_t=2.5$, $J_l=5$, $J_{c}=3$}
\label{fig4}
\end{figure}

A confirmation of this trend can be observed in Figs.(\ref{fig4}a-f) where $J_t = 2.5$. The 
rates of destabilisation of the excited states are higher than in the former cases as it can be seen 
in Figs.(\ref{fig4}c-d). This point is also reflected in the behaviour of the entropy $s$ 
which is larger than in the former case for $n = N$ and decreases more rapidly with decreasing $n$,
Fig.(\ref{fig4}f).\\  

\subsubsection{Third case: $L$= 8,  $J_t$=15, $J_l$=5, $J_{c}$=3}

For $L=8$ the Hilbert space is spanned by $N = 12870$ basis states with $M_{tot}=0$. The results 
are shown in  Figs.(\ref{fig6}a-f). The stability of the spectrum with decreasing space dimension
is stronger than the stability observed for $L=6$.  Indeed if $n/N$ defines the ratio of the number 
of states in the reduced space over the total number of states one finds $p(1) \sim 0.8 \%$ and 
$p(2) \sim 0.8 \%$ when $n/N \sim 0.07$ for $L=6$. For $L=8$ $p(1) \sim 0.8 \%$ when 
$n/N \sim 0.007 $ and $p(2) \sim 0.5 \%$ when $n/N \sim 0.02$. This shows a sizable improvement 
in the stability of the spectrum, at least in the specific domain where the coupling strength 
$J_t$ is large compared to the others.\\

The evolution of the spectrum and its stability with decreasing $J_t$  follows the same trend as 
in the case where $L = 6$.\\

\subsubsection{Remarks}

In Fig.(\ref{fig2}a) it is seen that the ground state shows "bunches" of energy fluctuations. The 
peaks are intermittent, they appear and disappear during the space dimension reduction process. They 
are small in the case where $J_t = 15$ but can grow with decreasing $J_t$ as it can be observed for 
$J_t = 2.5$. The subsequent stabilization of the ground state energy following such a bunch shows 
the effectiveness of the coupling constant renormalization which acts in a progressively reduced and 
hence incomplete basis of states.\\

\begin{figure}[ht]
\includegraphics{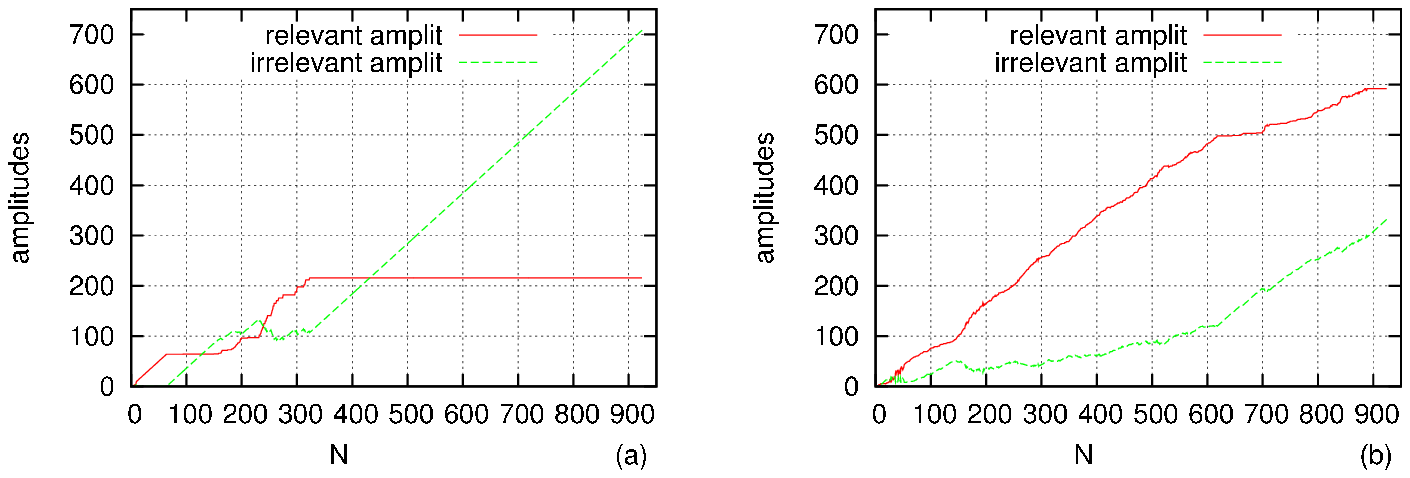}
\caption{$SU(2)-scheme$. $N$ is the dimension of the Hilbert space. $Amplitudes$ show the number of 
relevant -irrelevant amplitudes in the ground state eigenfunction. Relevant amplitudes are those for 
which $\{a_{1i}>\epsilon$, (here $\epsilon = 10^{-2}$), $i=1,\ldots,n\}$. The number of sites 
along a leg is $L=6$, (a) corresponds to $J_t=15$, (b) to $J_t=2.5$; $J_l=5$, $J_{c}=3$}
\label{fig5}
\end{figure}

These bunches of fluctuations are correlated with the change of the number of relevant amplitudes 
(i.e. amplitudes larger than some value $\epsilon$ as explained in the caption of Fig.(\ref{fig5})) 
during the reduction process.\\

\begin{figure}[ht]
\includegraphics{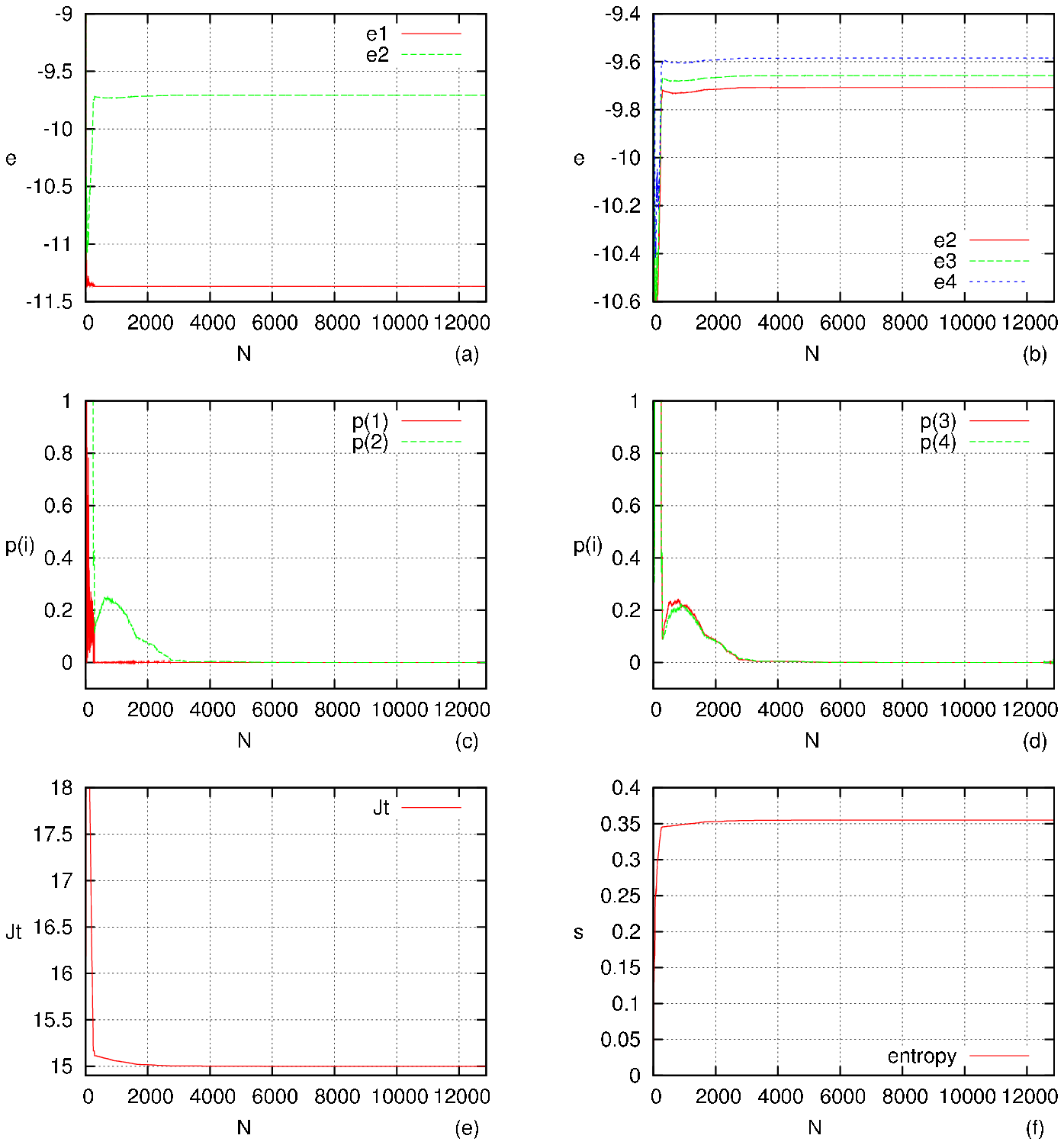}
\caption{$SU(2)- scheme$. $N$ is Hilbert space dimension. The  $\{e_i, i= 1,2,3,4\}$ are the 
energies per site. $L=8$ sites along a leg. $J_t=15$, $J_l=5$, $J_{c}=3$}
\label{fig6}
\end{figure}

Consider first the case where $J_t > J_l, J_c$. One notices in the caption of Fig.(\ref{fig5}a)
that up to $n\sim300$ the number of relevant amplitudes defined in Fig.(\ref{fig5}) stays stable 
like the ratios $\{p(i)\}$ in Figs.(\ref{fig2}c-d). For $158<n<300$ these ratios change quickly. 
A bunch of fluctuations appears in this domain of values of $n$ as seen in Figs.(\ref{fig2}c-d) 
and correspondingly the number of relevant amplitudes decreases steeply. For $60 < n \leq158$ 
the ratios $\{p(i)\}$ stay again stable as well as the number of relevant amplitudes. The $\{p(i)\}$ 
in Fig.(\ref{fig2}c-d) almost decrease back to their initial values. The same explanation is valid 
for $L=8$ in Fig.(\ref{fig6}). The analysis shows that these bunches of fluctuations 
signal the local elimination of relevant contributions of basis states to the physical states in 
the spectrum. The stabilization of the spectra which follows during the elimination process shows 
that renormalization is able to cure these effects.\\

In the case where $J_t < J_l, J_{c}$ shown in Fig.(\ref{fig5}b) the relevant and irrelevant
amplitudes move continuously during the reduction process and the corresponding $\{p(i)\}$
do no longer decrease to the values they showed before the appearance of the bunch of 
energy fluctuations as seen in Figs.(\ref{fig6}c-d). It signals the fact that the coupling 
renormalization is no longer able to compensate for the reduction of the Hilbert space 
dimensions.\\

\subsection{Spectra in the SO(4)-symmetry framework}

The reduction algorithm is now applied to the system described by the Hamiltonian $H^{(S,R)}$ 
given by Eq.~(\ref{eq12}) with a basis of states written in the $SO(4)$ symmetry framework. Like 
above we consider two cases corresponding to large and small values of $J_t$ relative 
to the strengths of the other coupling parameters.

\subsubsection{Reduction test for $L=6$, $J_t$ = 15, 2.5 and $J_l$=5, $J_{c}$=3}
 
\begin{figure}[ht]
\includegraphics{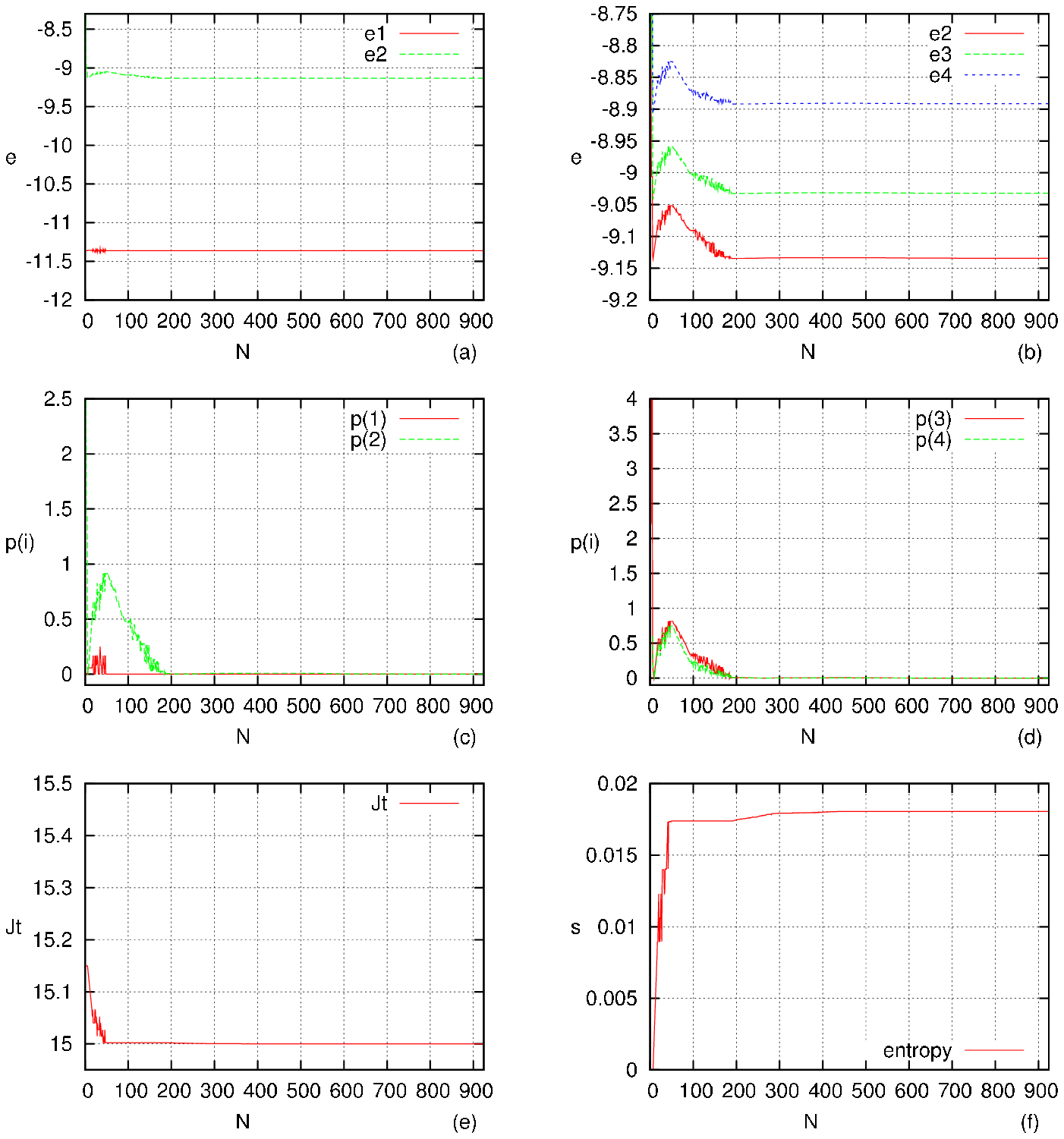}
\caption{$SO(4)- scheme$. $N$ is Hilbert space dimension. The  $\{e_i, i= 1,2,3,4\}$ are the 
energies per site. $L=6$ sites along the chain. $J_t=15$, $J_l=5$, $J_{c}=3$}
\label{fig7}
\end{figure}

Figs.(\ref{fig7}) show the behaviour of the spectrum for a system of size $L=6$. A large value of 
$J_t$, ($J_t = 15$), favours the dimer structure along rungs in the lowest energy state and 
stabilizes the spectrum down to small Hilbert space dimensions. This effect is clearly seen in 
Fig.(\ref{fig7}a), the ground state is very stable. The excited states are more affected, see  
Figs.(\ref{fig7}b), although they do not move significantly, Figs.(\ref{fig7}c-d). The 
renormalization of the coupling strength $J_t$ starts to work for $n \simeq 50$.\\

\begin{figure}[ht]
\includegraphics{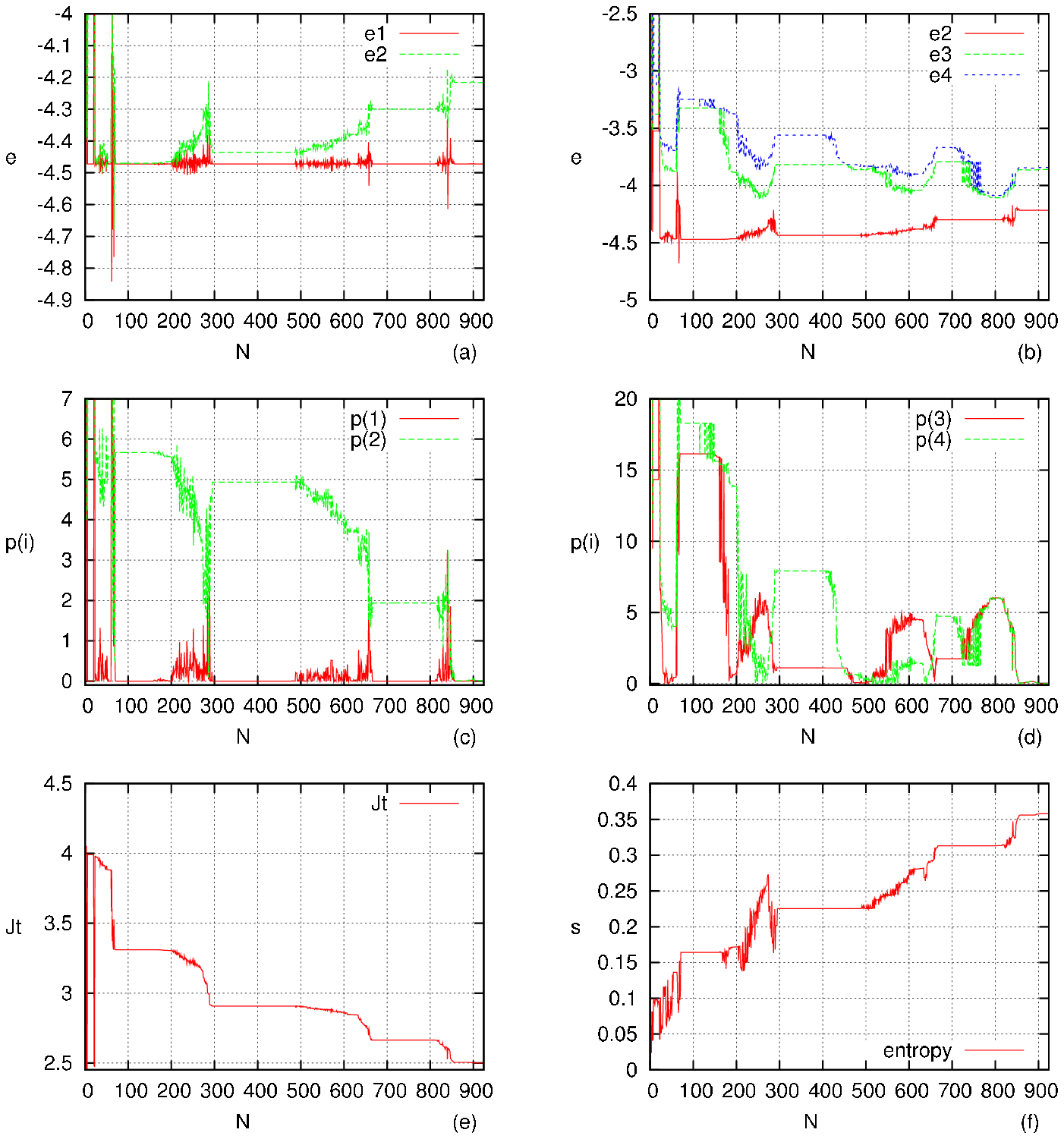}
\caption{$SO(4)- scheme$. N is Hilbert space dimension.The  $\{e_i, i= 1,2,3,4\}$ are the energies 
per site. $L=6$ sites along the chain. $J_t=2.5$, $J_l=5$, $J_{c}=3$}
\label{fig8}
\end{figure}

The situation changes progressively with decreasing values of $J_t$. Figs.(\ref{fig8}) show the 
case where $J_t=2.5$. The ground state energy experiences sizable bunches of fluctuations 
like in the $SU(2)$ scheme, but much stronger than in this last case. The same is true for the 
excited states which is reflected through all the quantities shown in Figs.(\ref{fig8}), in 
particular $J_t$, Fig.(\ref{fig8}e). The arguments used in the $SU(2)$-scheme about relevant and
irrelevant amplitudes are also valid here.\\ 

The result shows that the renormalization procedure is quite sensitive to the symmetry scheme chosen 
in Hilbert space. It is expected that essential components of the ground state wavefunction get 
eliminated early during the process when the rung coupling gets of the order of magnitude or smaller 
than the other coupling strengths.\\
 
\subsection{Summary}

The present results lead to two correlated remarks. The efficiency of the algorithm is different
in different sectors of the coupling parameter space. In the case of the frustrated ladders 
considered here the algorithm is the more efficient the stronger the coupling between rung sites 
$J_t$. Second, this behaviour is strongly related to the symmetry representation in which the basis 
of states is defined. The $SU(2)$ representation leads to a structure of the wavefunctions (i.e. 
the size of the amplitudes of the basis states) which is very different from the one obtained in 
the $SO(4)$ representation. For large values of $J_t$ the spectrum is more stable in the $SO(4)$ 
scheme. For small values of $J_t$ the stability is better realized in the $SU(2)$ scheme. Finally, 
in the regime where $J_t > J_l, J_{c}$, one observes that the reduction procedure is the more 
efficient the closer $J_l$ to $J_{c}$. This effect can be understood and related to previous 
analytical work in the $SO(4)$ framework~\cite{jr2}.\\
 
\section{Conclusions and outlook.}

In the present work we tested and analysed the outcome of an algorithm which aims to reduce the 
dimensions of the Hilbert space of states describing strongly interacting systems. The reduction is 
compensated by the renormalization of the coupling strengths which enter the Hamiltonians of the 
systems. By construction the algorithm works in any space dimension and may be applied to the study 
of any microscopic $N$-body quantum system. The robustness of the algorithm has been applied to 
frustrated quantum spin ladders.\\

The analysis of the numerical results obtained in applications to quantum spin ladders leads to the 
following conclusions.  

\begin{itemize} 

\item The stability of the low-lying states of the spectrum in the course of the reduction procedure 
depends on the relative values of the  coupling strengths. The ladder favours a dimer structure 
along the rungs, i.e. stability is the better the larger the transverse coupling strength $J_t$.\\  

\item The evolution of the spectrum depends on the initial size of Hilbert space. The larger the 
initial space the larger the ratio between the initial number of states and the number of states 
corresponding to the limit of stability of the spectrum.\\ 

\item The efficiency of the reduction procedure depends on the symmetry frame in which the basis of 
states is defined. It appears clearly that the evolution of the spectrum described in an $SU(2)$ 
scheme is significantly different from the evolution in an $SO(4)$ scheme. This is again 
understandable since different symmetry schemes partition Hilbert space in different ways and favour 
one or the other symmetry depending on the relative strengths of the coupling constants. 

Local spectral instabilities appearing in the course of the reduction procedure are correlated 
with the elimination of basis states with sizable amplitudes in the ground state wavefunction.
One or another representation can be more efficient for a given set of coupling parameters because 
it leads to physical states in which the weight on the basis states is concentrated in a different 
number of components. This point is strongly related to the correlation between quantum entanglement 
and symmetry properties which are presently under intensive scrutiny, see f.i. ~\cite{kor} and 
refs. therein.  
  
\end{itemize} 

Further points are worthwhile to be investigated:\\

\begin{itemize}
 
\item In the present approach the sequential reduction of space dimensions followed an energy 
criterion. It might be judicious to classify the sequence of states to be eliminated starting 
with those which have the smallest amplitude in the ground state wavefunction. The two procedures 
should however be correlated if not equivalent.\\ 

\item We expect to extend the study to systems of higher space dimensions, f.i. $2d$.

\item The present approach relies on an algorithm which is able to recognize the existence of first 
and higher order critical points  ~\cite{khri}. It is of interest to apply the algorithm in the 
neighbourhood of such points. Its behaviour could help to identify them. Work on this point is
under way.\\   

\item The algorithm can be extended to systems at finite temperature ~\cite{jr} and more than one 
coupling constant renormalization.

\end{itemize}

The authors would like to thank Dr. A. Honecker for interesting discussions. One of us (T.K.) 
acknowledges the help of Drs. E. Caurier and F. Nowacki on technical aspects concerning the 
implementation of the Lanczos algorithm.

\end{document}